\documentclass[twocolumn,showpacs,preprintnumbers,amsmath,amssymb]{revtex4}
\usepackage[dvips]{graphicx}
\usepackage{dcolumn}

\begin{document}

\title
{Impact of Strong Anisotropy on Phase Diagram of Superfluid $^3$He in Aerogels
}

\author{Tomohiro Hisamitsu, Masaki Tange, and Ryusuke Ikeda}

\affiliation{Department of Physics, Graduate School of Science, Kyoto University, Kyoto 606-8502, Japan
}

\date{\today}

\begin{abstract} 
Recently, one analog of the Anderson's Theorem for the $s$-wave superconductor has attracted much interest in the context of the $p$-wave polar pairing state of superfluid $^3$He in a model aerogel in the limit of strong uniaxial anisotropy. We discuss to what extent the theorem is satisfied in the polar phase in real aerogels by examining the normal to polar transition temperature $T_c$ and the low temperature behavior of the superfluid energy gap under an anisotropy of a moderate strength and comparing the obtained results with experimental data. The situation in which the Anderson's theorem clearly breaks down is also discussed. 
\end{abstract}


%\keyword{}

\maketitle

%\section{Introduction}
Recent observations on superfluid $^3$He in anisotropic aerogels have clarified profound roles of an anisotropy for the superfluid phase diagram and properties. The polar pairing state \cite{AI06} has been discovered in nematic aerogels with a nearly one-dimensional structure \cite{Dmitriev1}. It has been found that this polar pairing state does not occur when the magnetic scattering effect due to the solid $^3$He localized on the surface of the aerogel structure is active \cite{Dmitriev2}. This high sensitivity to the type of "impurity" scatterings of the superfluid phase diagram is not easily explained within the original theoretical model assuming a {\it weak} global anisotropy of the aerogel structure 
\cite{AI06,Sauls}. 

It has been pointed out that, when the aerogel structure is in the limit of strong anisotropy so that the scattering is specular along the anisotropy axis, the normal to polar transition temperature $T_c(P)$ should be insensitive to the (nonmagnetic) impurity scattering strength \cite{Fomin18}. 
%Effects of such columnar or line-like scattering potentials on the bosonic 
%degrees of freedom have been previously considered in other contexts 
%\cite{NV,RI99}. 
Recently, this result analogous to the Anderson's Theorem in the $s$-wave superconductor \cite{Anderson} has attracted much interest \cite{Eltsov,Dmitriev3} in relation to the low temperature behavior of the energy gap in the polar phase and to the robustness of the $p$-wave superfluid polar phase in relatively dense nematic aerogels. 
Previously, various features seen in superfluid $^3$He in nematic aerogels \cite{Dmitriev1,Dmitriev3} have been discussed based on the model assuming the {\it weak} anisotropy \cite{AI06}. Here, the anisotropy is measured by the size of the correlation length $L_z$ of the random scattering potential. Once taking account of the puzzling result \cite{Dmitriev2} brought by the magnetic impurities altogether, the approach starting from the side of the strong anisotropy may be more appropriate. Further, the polar phase has been detected so far only in the nematic aerogels. Then, one might wonder whether the polar phase occurs only in the limit of strong anisotropy. However, $L_z$ in nematic aerogels seems to be finite if taking account of the fact that splayed strands and crossings between straight strands are seen in real images of the nematic aerogels \cite{Dmitriev1,Eltsov}. 

In this communication, consequences of the strong anisotropy in the phase diagram of superfluid $^3$He in aerogels with no magnetic scattering effect are studied in details. It is found that, in the weak-coupling BCS approximation, the impurity-scattering independent $T_c$ is approximately satisfied even in the scattering potential model with a finite correlation length $L_z$ along the anisotropy axis, implying that the Anderson's Theorem is apparently satisfied over a wide range of the strengths of the anisotropy. Thus, we argue that, consistently with the original argument \cite{AI06}, the polar phase may be realized in aerogels with a global anisotropy of a moderate magnitude. Further, the dependences of the superfluid energy gap $|\Delta(T)|$ on the strengths of the impurity scattering and the anisotropy are also examined, and the $T^3$ behavior arising from the horizontal line node of $|\Delta(T)|$ in the polar pairing symmetry is found to be robust against changes of the impurity strength and the anisotropy. 
%the the corresponding A phase does not occur as far as the strong-coupling (SC)% effect is neglected. As pointed out in Ref.\cite{Fomin18}, 
Further, the situation in which $T_c(P)$ is also reduced so that the Anderson's Theorem is not satisfied will also be discussed. 

First, let us describe how the Anderson's Theorem occurs in the context of the $p$-wave superfluid phase in an environment with nonmagnetic elastic impurity scatterings. The starting model of our analysis to be performed below is the BCS Hamiltonian for a spatially uniform equal-spin paired state in zero magnetic field 
\begin{eqnarray}
{\cal H}_{\rm BCS} \!\!&-&\!\! \mu N = \sum_{{\bf p}, \sigma} \biggl[\xi_{\bf p} a^\dagger_{{\bf p}, \sigma} a_{{\bf p}, \sigma} - \frac{1}{2} ( \Delta_{\bf p}^* a_{{\bf p}, \sigma} a_{-{\bf p}, \sigma} + {\rm h.c.}) \biggr] \nonumber \\ 
&+& g^{-1} V |\Delta|^2,  
\end{eqnarray}
where $g$ is the strength of the attractive interaction, $V$ is the system volume, $|\Delta|$ is the maximum of the quasiparticle energy gap, and $\xi_{\bf p}$ is the quasiparticle energy measured from the Fermi energy $\mu$. 

The total Hamiltonian ${\cal H}$ is the sum of eq.(1) and the nonmagnetic impurity potential term 
\begin{equation}
{\cal H}_{\rm imp} = \int d^3{\bf r} \, u({\bf r}) \, n({\bf r}),  
\end{equation}
where $n({\bf r})$ is the particle density operator. As usual, the impurity scattering can be modelled by the 
correlator 
\begin{equation}
W({\bf r}) = 2 \pi N(0) \tau \, \langle u({\bf r}) u(0) \rangle_{\rm imp}, 
\end{equation}
or its Fourier transform $w({\bf k}) = \int d^3{\bf r} W({\bf r}) e^{{\rm i}{\bf k}\cdot{\bf r}}$, with $\langle u \rangle_{\rm imp}=0$, where $\langle \,\,\, \rangle_{\rm imp}$ denotes the random average, $N(0)$ is the density of states on the Fermi surface per spin in the normal state, and $\tau$ is the relaxation time of the normal quasiparticle in the case with no anisotropy. For simplicity, the Born approximation will be used to incorporate the impurity-scattering effect in the Green's functions for the quasiparticles in an equal-spin paired superfluid state. Then, we have a mean field problem for spin-less Fermions, and solving the corresponding gap equation can be performed in quite the same manner as in the $s$-wave paired case \cite{AGD}. 
%Details of the analysis will be given in   . 
The resulting gap equation can be expressed in the form 
%\begin{widetext}
\begin{equation}
{\rm ln}\biggl(\frac{T}{T_{c0}(P)} \biggr) = \pi T \sum_{\varepsilon} \biggl[ \frac{-1}{|\varepsilon|} + 3 \biggl\langle  \frac{ \Delta_{\bf p} {\tilde \Delta}_{\bf p} \Delta^{-2}}{\sqrt{{\tilde \varepsilon}_{\bf p}^2 + |{\tilde \Delta}_{\bf p}|^2}} \biggr\rangle_{{\hat p}} \biggr], 
\end{equation}
%\end{widetext}
where $\varepsilon= \pi T (2m+1)$ with integer $m$, $T_{c0}(P)$ is the superfluid transition temperature of the bulk liquid, and $\langle \,\,\, \rangle_{{\hat p}}$ denotes the angle average on the unit vector ${\hat p}$ over the Fermi surface. Further, in eq.(4), 
\begin{eqnarray}
i{\tilde \varepsilon}_{\bf p} &=& i\varepsilon - \frac{1}{2 \pi N(0) \tau} \int_{\bf q} w({\hat p} - {\hat q}) {\cal G}_{\bf q}(\varepsilon), \nonumber \\
{\tilde \Delta}_{\bf p} &=& \Delta_{\bf p} - \frac{1}{2 \pi N(0) \tau} \int_{\bf q} w({\hat p} - {\hat q}) [{\cal F}^\dagger_{\bf q}(\varepsilon)]^* ,
\end{eqnarray}
and 
\begin{eqnarray}
{\cal G}_{\bf p}(\varepsilon) &=& \frac{-{\rm i}{\tilde \varepsilon}_{\bf p} - \xi_{\bf p}}{{\tilde \varepsilon}_{\bf p}^2 + \xi_{\bf p}^2 + |{\tilde \Delta}_{\bf p}|^2}, \nonumber \\
{\cal F}^\dagger_{\bf p}(\varepsilon) &=& - \frac{ {\tilde \Delta}_{\bf p}^*}{
{\tilde \varepsilon}_{\bf p}^2 + \xi_{\bf p}^2 + |{\tilde \Delta}_{\bf p}|^2} 
\end{eqnarray}
are the impurity-averaged Matsubara Green's functions \cite{AGD}. 
%averaged Matsubara Green's functions, the normal component ${\cal G}_{\bf p}(\v%arepsilon)$ and the anomalous one ${\cal F}^\dagger_{\bf p}(\varepsilon)$, are %given by  averaged over the impurity scattering effect 

As a model of the impurity correlator (3) in the presence of a stretched anisotropy favoring the polar phase in which $\Delta_{\bf p} = \Delta {\hat p}_z$, we will use the following expression 
\begin{eqnarray}
W({\bf r}) &=& \frac{k_{\rm F}}{2} \, \delta^{(2)}({\bf r}_\perp) \,  \exp(-|z|/L_z) \nonumber \\
&\times& [ 1 + \theta(1 - |\delta_u|) (|\delta_u|^{-1/2} - 1) ], 
\end{eqnarray}
where $L_z$ is the correlation length defined along the anisotropy axis on the random distribution of the potential $u({\bf r})$, and the $z$-axis is chosen here as the anisotropy axis or stretched direction. The size of the anisotropy is measured by $|\delta_u|=k_{\rm F}^2 L_z^2$, while the measure of the impurity strength is $1/(\tau T_{c0})$ \cite{Thuneberg98}, where $k_{\rm F}$ is the Fermi wave number. Then, the Fourier transform $w({\bf k})$ of $W({\bf r})$ becomes 
\begin{equation}
w({\hat {\bf k}}) = \frac{\sqrt{|\delta_u|}}{1+|\delta_u| {\hat k}_z^2} \, \biggl( 1 + (|\delta_u|^{-1/2} - 1) \theta(1 - |\delta_u|) \biggr). 
\end{equation}

Equation (8) has the following limiting cases. For the weak anisotropy, $|\delta_u| < 1$, this model reduces to the expression, $w({\hat {\bf k}}) \simeq 1 - |\delta_u| {\hat k}_z^2$, introduced in Ref.\cite{AI06}. 
%%%%%%%%%%%%%%%%%%%
\begin{figure}[b]
{\includegraphics[scale=1.2]{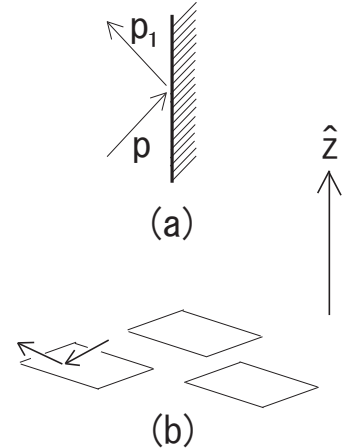}}
\caption{(Color online) (a) Specular reflection expressing the conservation of the component along the ${\hat z}$-axis (vertical direction) of the momentum. 
(b) Rough picture of planar scattering centers positioned randomly and with their surface perpendicular to the $z$-axis. On each center, the component perpendicular to the $z$-axis (i.e., parallel to the surface) of the momentum 
is conserved at each scattering event. 
 } 
\label{fig.1}
\end{figure}
%%%%%%%%%%%%%%%%%
The opposite limit of the infinite $|\delta_u|$ corresponds to the case with the impurity scattering persistent along the stretched direction. In this case, $w({\hat {\bf k}})$ reduces to 
\begin{equation}
w_\infty({\hat k}) = \pi k_{\rm F} \delta(k_z),   
\end{equation}
implying that, as sketched in Fig.1(a), the scattering is specular along the $z$-axis. The present model (7) interpolating the above-mentioned two limits has been used to study how the half-quantum vortex (HQV) pair, which should appear in the polar phase, survives in the PdB phase at lower temperatures \cite{TI}. For any value of the anisotropy $\delta_u$ and the impurity strength $1/(\tau T_{c0})$, the polar to normal transition temperature $T_c(P)$ and the superfluid gap $|\Delta(T)|$ in the polar phase can be numerically obtained using eqs.(4) and (8). 

Now, it is easy to verify the Anderson's Theorem for the polar pairing state with $\Delta_{\bf p} = \Delta {\hat p}_z$ in the limit of the strong anisotropy. In fact, by applying eq.(9) to eq.(5), any $1/(\tau T_{c0})$ dependence in the last term of eq.(4) is cancelled between the denominator and numerator of the term, and, as in the $s$-wave pairing case, the gap equation (4) becomes its expression in clean limit or for the bulk liquid. 

%%%%%%%%%%%%%%%%%%%
\begin{figure}[b]
{\includegraphics[scale=1.6]{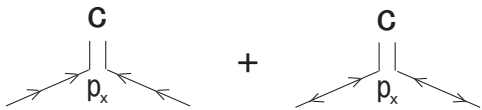}}
\caption{(Color online) Diagrams expressing the gap equation linearized with respect to the order parameter of the PdB phase. The parameter $c$ means the order parameter of the PdB phase, and each vertex carries a component perpendicular to the $z$-axis of the momentum. The impurity-averaged Green's functions ${\cal G}$ and ${\cal F}^\dagger$ in eq.(6) are indicated by a line with a double arrow and that with a left right arrow, respectively. 
 } 
\label{fig.2}
\end{figure}
%%%%%%%%%%%%%%%%%

Next, let us determine the polar to PdB transition temperature 
$T_{\rm PB}(P)$. In the present weak-coupling approach, the polar-distorted A (PdA) phase \cite{Dmitriev1,AI06,Eltsov} does not appear, and, on cooling, the polar phase is transformed into the PdB phase through a continuous transition. Since the real polar to PdA transition is also continuous \cite{Dmitriev1,AI06}, however, the $T_{\rm PB}$ line obtained here is expected to be qualitatively comparable with the polar to PdA transition line. The $T_{\rm PB}(P)$ line is easily obtained according to the diagrams sketched in Fig.2 representing the gap equation linearized with respect to the order parameter of PdB state by using the quantities characterizing the polar pairing state. Then, $T_{\rm BP}$ is given by the temperature $T$ satisfying 
\begin{equation}
{\rm ln}\biggl(\frac{T}{T_{c0}(P)} \biggr) = \pi T \sum_{\varepsilon} \biggl[ \frac{-1}{|\varepsilon|} + \frac{3}{2} \biggl\langle  \frac{1-{\hat p}_z^2}{\sqrt{{\tilde \varepsilon}_{\bf p}^2 + |{\tilde \Delta}_{\bf p}|^2}} \biggr\rangle_{{\hat p}} \biggr].
\end{equation} 

%%%%%%%%%%%%%%%%%%%
\begin{figure}[t]
{\includegraphics[scale=0.7]{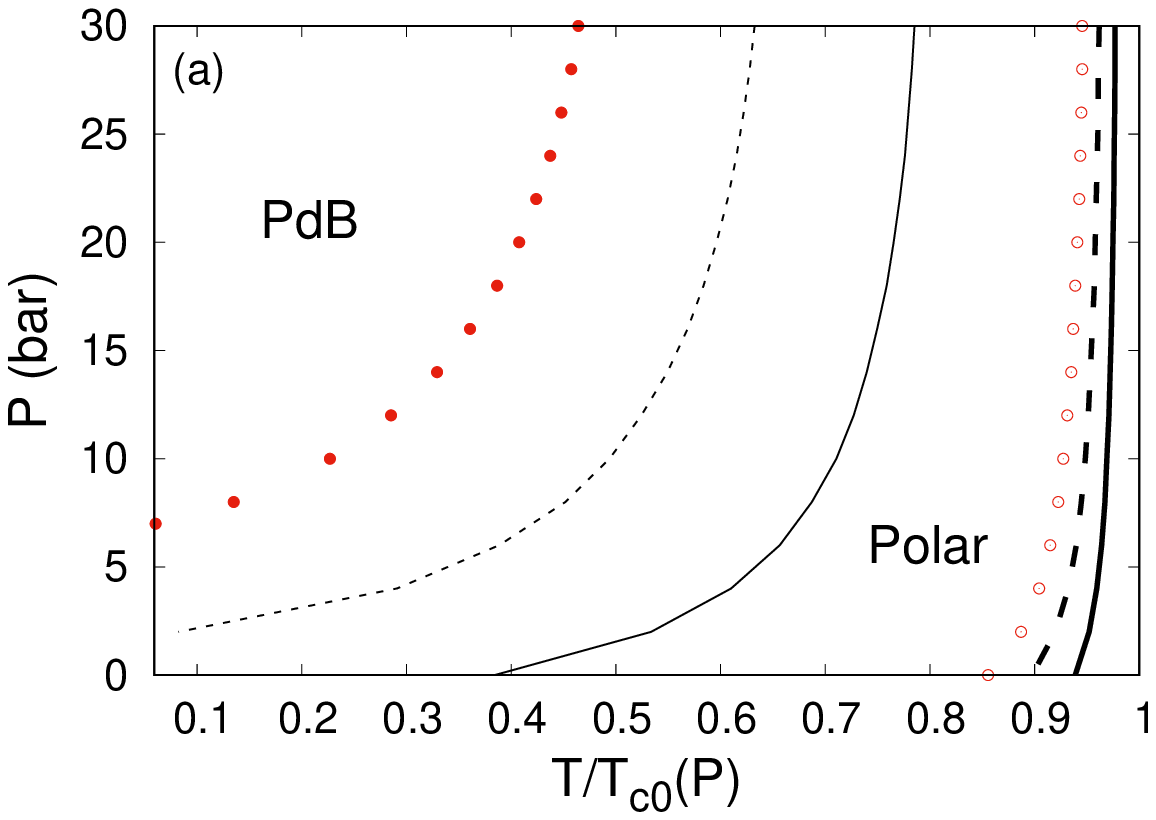}}
{\includegraphics[scale=0.7]{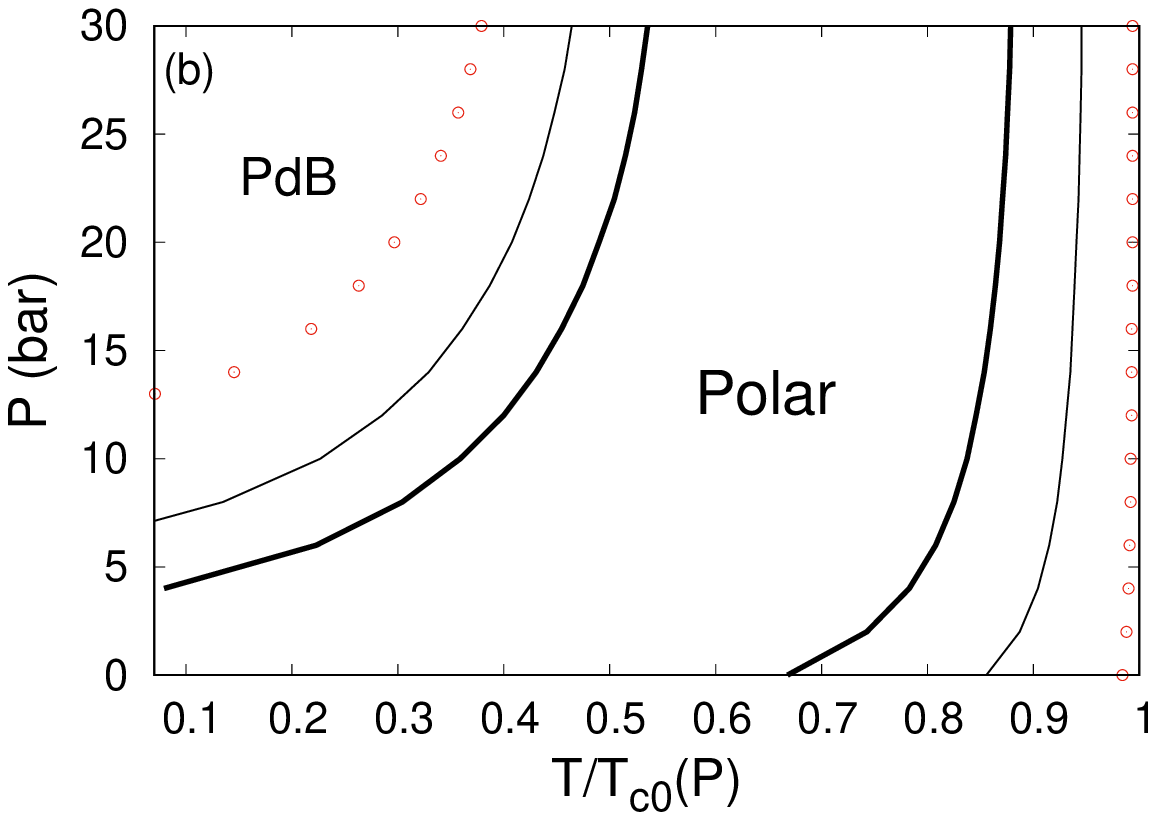}}
\caption{(Color online) (a) Pressure ($P$) v.s. temperature ($T$) superfluid phase diagram obtained by changing the parameter $\tau^{-1}$ measuring the impurity strength under a fixed magnitude of the anisotropy $|\delta_u|=30$, i.e., $k_{\rm F} L_z=5.48$. The temperatures $T_c(P)$ and $T_{\rm PB}(P)$ are represented by the thick and thin solid lines for $(2 \pi \tau)^{-1}({\rm mK}) = 0.3$, the thick and thin dashed lines for $(2 \pi \tau)^{-1} = 0.5$, and the open and closed circles for $(2 \pi \tau)^{-1} = 0.7$, respectively. Note that the variable of the horizontal axis is $T/T_{c0}(P)$, i.e., the temperature normalized by the bulk superfluid transition temperature $T_{c0}$ at each $P$. (b) Corresponding ones obtained by changing the anisotropy or the correlation length $L_z$ under the fixed impurity strength $(2 \pi \tau)^{-1}({\rm mK}) = 0.7$. The temperatures $T_c(P)$ and $T_{\rm PB}(P)$ are represented by the right and left thick solid curves for $|\delta_u|=4.4$ and the right and left lines expressed in terms of the open circles for $|\delta_u|=3 \times 10^3$, respectively. The right and left thin solid curves are the open and closed circles in (a), respectively. 
 } 
\label{fig.3}
\end{figure}
%%%%%%%%%%%%%%%%%
Examples of the $T_c(P)$ and $T_{\rm PB}(P)$ obtained numerically from eqs.(4) and (10) are presented in Fig.3, where the experimental data on $T_{c0}(P)$ \cite{VW} were used. As is seen in Fig.3(a) where a moderately large anisotropy $|\delta_u|=30$ is used, $T_c(P)$ weakly depends on the impurity strength $\tau^{-1}$. In general, for a stronger anisotropy, the $\tau^{-1}$-dependence of $T_c$ becomes weaker, while the corresponding one of $T_{\rm PB}$ becomes stronger. At higher pressures, the pressure dependence of $T_c/T_{c0}(P)$ is quite weak, reflecting the proximity to the limit of strong anisotropy in which the Anderson's Theorem is exact, while $T_c/T_{c0}$ is lowered at low enough $P$-values because of an increase of the dimensionless impurity strength $1/(\tau T_{c0}(P))$. In contrast to $T_c$, however, $T_{\rm PB}$ is quite sensitive to the impurity strength and rapidly decreases with increasing $1/(\tau T_{c0})$. Thus, the temperature range of the polar phase is wider for a lower $P$. 

Further, as Fig.3 (b) shows, an increase of the anisotropy extends the region of the polar phase: With increasing the anisotropy $|\delta_u|$, $T_c$ is increased and approaches $T_{c0}$, while $T_{\rm PB}$ decreases and approaches its {\it finite} value in the limit of strong anisotropy (see Fig.3 (b)). In any case, the temperature range of the polar phase at a fixed $P$ becomes wider with increasing the anisotropy and/or the impurity strength. 

The results on $T_c(P)$ in Fig.3 will be compared with the corresponding curves determined in experiments \cite{Dmitriev1,Dmitriev2,Dmitriev3,Dmitriev4}. In aerogels, an effective decrease of the porosity leads to an enhancement of the "impurity" scattering via the aerogel structure \cite{Dmitriev2}. In fact, Fig.1(a) and Fig.2(a) and (c) in Ref.\cite{Dmitriev4} have shown a slight decrease of $T_c$ and a drastic decrease of the transition line to the PdA phase due to a reduction of the porosity. This tendency of the two transition curves is consistent with the features seen with increasing $1/(\tau T_{c0})$ in Fig.3(a). By combining this observation with the $T_c$ curve under the large enough anisotropy, $|\delta_u|=3 \times 10^3$ in Fig.3(b), it is natural to regard the nematic aerogels in which the polar phase of superfluid $^3$He is realized as random media with a {\it finite} correlation length of the scattering potential. 
%Without including the strong coupling effects which is indispensable of 
%appearance of the PdA phase, this identification will be possible because the 
%polar to PdA transition is also continuous \cite{AI06}. 
Nevertheless, examining the superfluid polar phase in the nematic aerogels by starting from the limit of strong anisotropy where the Anderson's Theorem is satisfied is a proper description. 
%%%%%%%%%%%%%%%%%%%
\begin{figure}[t]
{\includegraphics[scale=0.7]{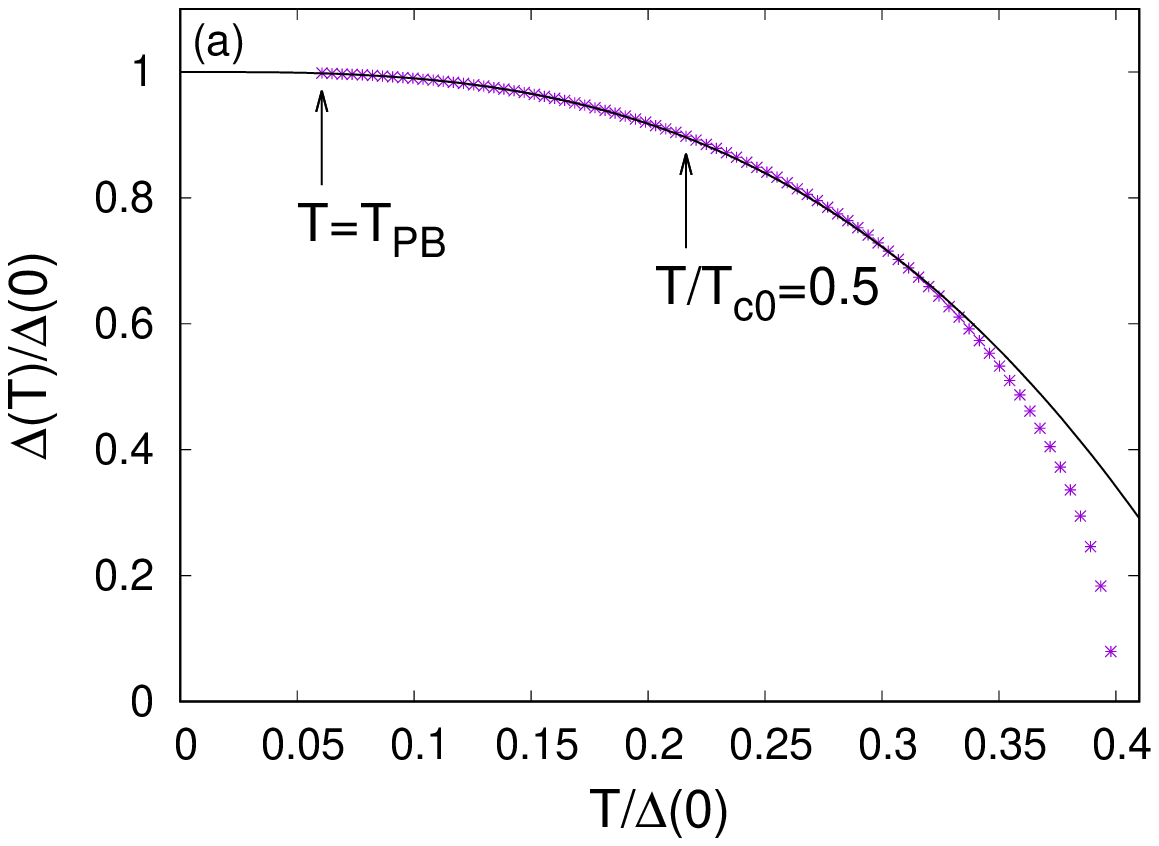}}
{\includegraphics[scale=0.7]{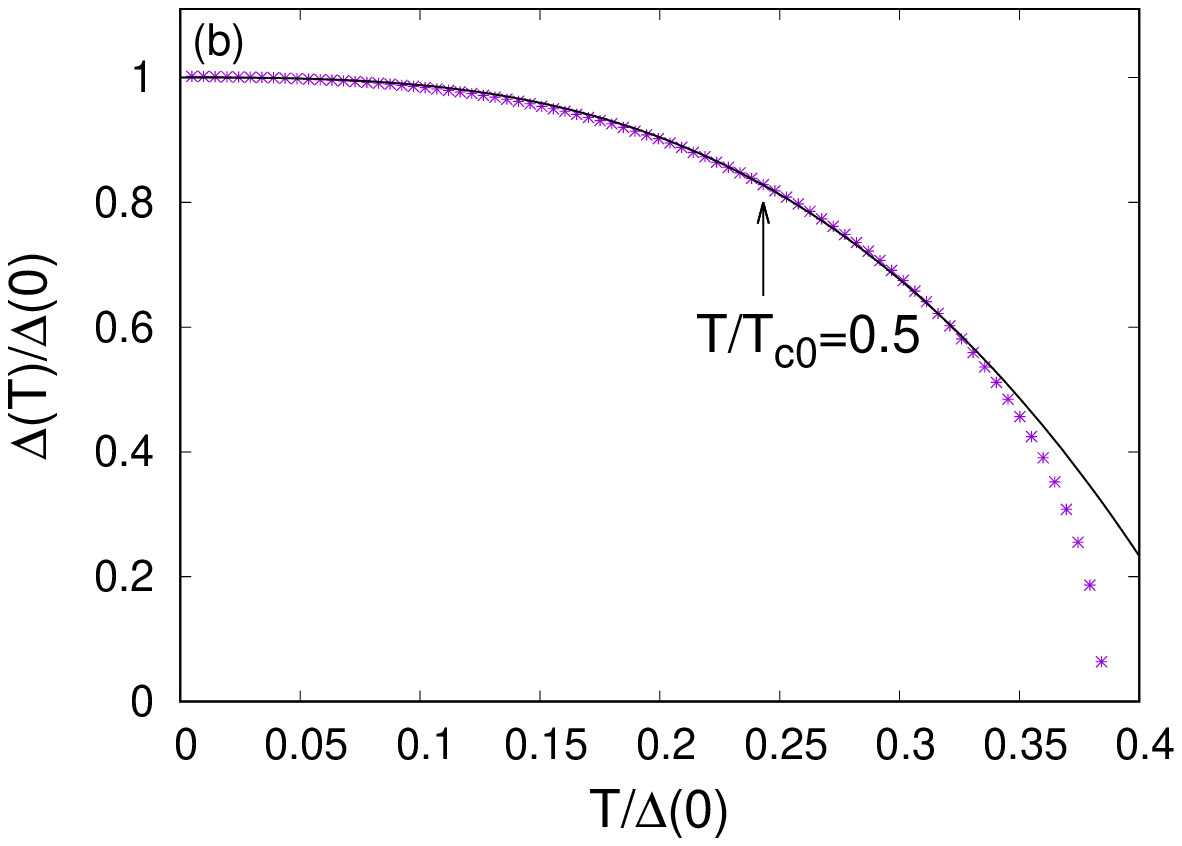}}
\caption{(Color online) Temperature dependence of $|\Delta|$ for the parameter values $|\delta_u|=30$ and $(2 \pi \tau)^{-1}=1({\rm mK})$ (a) at 30(bar) and (b) at 0(bar). In both cases, the $T^3$ behavior is nicely seen at least in the range $T < 0.65 T_c$ (Note that, in the figures, the temperature is represented in unit of $|\Delta(0)|$ in each case). The thick solid curve expresses eq.(11), where the ${\overline a}$ value is $10.1$ in (a) and $12.0$ in (b). 
 } 
\label{fig.4}
\end{figure}
%%%%%%%%%%%%%%%%% 

Next, as another quantity related to the Anderson's Theorem, let us examine the temperature dependence of the energy gap $|\Delta(T)|$ of quasiparticles in the polar phase. As indicated elsewhere \cite{Eltsov}, the energy gap difference $|\Delta(0)|-|\Delta(T)|$ estimated from the NMR frequency data in the polar phase at 30 (bar) is proportional to $T^3$, reflecting the presence of a line node in $|\Delta(T)|$. Since the relevant energy scale at low $T$ is not $T_c$ but $|\Delta(0)|$, we will express the $T^3$ behavior in the 
form 
\begin{equation}
1 - \frac{|\Delta(T)|}{|\Delta(0)|} = {\overline a} \frac{T^3}{|\Delta(0)|^3}. 
\end{equation}
This relation to be satisfied in the polar phase in aerogels would indicate that, irrespective of the presence of the impurity scattering effect, the line node of $|\Delta(T)|$ in the polar phase remains well defined. 
According to the {\it calculation} \cite{Eltsov} in the weak-coupling approximation and clean limit, the coefficient ${\overline a}$ takes the value $8.49$, while the estimated ${\overline a}$-value taken from NMR data in a nematic aerogel at 30 (bar) was $0.38 (\Delta(0)/T_c)^3$ \cite{Eltsov}. According to Ref.\cite{Eltsov}, this estimated coefficient may become comparable \cite{Eltsov} with the weak coupling value $8.49$ in the limit of strong anisotropy if the strong coupling effect \cite{VW} enhancing $|\Delta(0)|$ is taken into account. In a strongly anisotropic case, $|\delta_u|=3 \times 10^3$, we have obtained the value ${\overline a}=8.97$ comparable with the weak coupling value mentioned above. 

On the other hand, as presented in Fig.4, our results for the moderately strong anisotropy, $|\delta_u|=30$, clearly show an effect of the impurity scattering on the coefficient of the $T^3$ term. Here, a stronger scattering strength $(2 \pi \tau)^{-1}=1({\rm mK})$ than those used in Fig.3 was used to obtain a wider polar region at lower temperatures. Although the $T^3$ behavior is still well defined in $T < 0.65 T_{c0}$ irrespective of the pressure value, the coefficient ${\overline a}$ is, as mentioned in Fig.4's caption, enhanced especially at lower pressures. If the strong coupling effect is taken into account, according to Ref.\cite{Eltsov} the coefficient $a \equiv {\overline a}(T_c/\Delta(0))^3$ at 30 (bar) would remarkably decrease so that the estimated value $a=0.38$ \cite{Eltsov} may be explained. At zero pressure, however, the strong coupling effect is not effective so that the coefficient $a$ of the $T^3$ behavior at lower pressures may show a large value of order unity. Examining the $T^3$ term of the energy gap at low pressures may become a test for the present theory. 

It is valuable to point out that the $p$-wave Anderson's Theorem on the superfluid transition temperature is also satisfied in the case of a normal to (distorted) A phase transition under plane-like defects with no two-dimensional momentum transfer (See Fig.1 (b)) if the ${\bf l}$-vector of this A phase is oriented along the normal of the plane of the defects. In fact, when the bare pairing vertex is $p_k (\delta_{j,k} - {\hat z}_j {\hat z}_k)$, and eq.(9) is replaced by the form proportional to $\delta({\hat k}_x) \delta({\hat k}_y)$, the superfluid transition temperature resulting from eq.(4) becomes $T_{c0}$ irrespective of the strength of the impurity scattering. In principle, such a situation can be realized in planar aerogels and would result in an extension of the temperature width of the planar-distorted A phase region at {\it lower} pressures and hence, according to Ref.\cite{NI}, in a realization of HQVs in the chiral A phase. 

As is well known in the context of the dirty $s$-wave superconductors, the Anderson's Theorem breaks down in systems with a strong enough impurity scattering due to the impurity effect in the repulsive channels of the quasiparticle interaction \cite{Finkelstein}. In fact, the $T_c(P)$ curve reported in Fig.4 of Ref.\cite{Dmitriev2} shows a remarkable deviation from $T_{c0}(P)$. Further, it is possible that the $\tau^{-1}$ dependence of $T_c(P)$ obtained under a finite anisotropy of the type seen in Fig.3(a) occurs even in the limit of strong anisotropy due to the above-mentioned mechanism associated with the Anderson localization, because $1/(\tau T_{c0}(P))$ is the pressure-dependent strength of the impurity 
scattering. To clarify to what extent this Anderson localization effect is effective in real systems, further comparison between the theoretical results and new data will be necessary in future. 

In conclusion, we have investigated to what extent the Anderson's theorem is satisfied in the polar phase by assuming the correlation length of the random potential in nematic aerogels to be long but finite. It has been found that the low temperature behavior of the superfluid energy gap stemming from the presence of the horizontal line node is robust against the impurity scattering and that the resulting phase diagram is qualitatively consistent with the available experimental data.

One of the authors (R.I.) is grateful to Vladimir Dmitriev and Bill Halperin for useful discussions. The present work was supported by JSPS KAKENHI (Grant No.16K05444).

\end{document}